\documentclass[onecolumn,showpacs,preprint,preprintnumbers,amsmath,amssymb]{revtex4}
\usepackage{amssymb}
\usepackage[dvips]{graphicx}
\begin{document}

\title{ Unconventional pairs glued  by conventional phonons in cuprate superconductors.}
\author{A. S. Alexandrov}

\affiliation{Department of Physics, Loughborough University,
Loughborough LE11 3TU, United Kingdom\\}

\begin{abstract}
It has
gone almost unquestioned that   superexchange  in the $t-J$ (or  Hubbard) model, and
 not phonons,  is responsible for the unconventional ("d-wave") pairing symmetry of cuprate superconductors. However a number of advanced
  numerical  studies   have not found
superconductivity in the Hubbard (or $t-J$) model. On the other hand compelling experimental evidence for a strong electron-phonon interaction (EPI)
has currently arrived. Here I briefly review some phonon-mediated unconventional pairing mechanisms. In particular the
anisotropy of  sound velocity  makes the  phonon-mediated
attraction of electrons non-local in space providing unconventional
Cooper pairs  with a nonzero orbital momentum already in the framework of the conventional BCS theory with  weak EPI. In the opposite limit of strong EPI
 rotational symmetry breaking  appears as a result of a reduced Coulomb repulsion between unconventional
bipolarons. Using the variational Monte-Carlo method we have found that a relatively
weak finite-range  EPI induces a d-wave BCS
 state also in doped Mott-Hubbard insulators or
strongly-correlated metals.    These results tell us that poorly screened EPI with
conventional  phonons  is responsible for the unconventional pairing in cuprate superconductors.

\vspace{0.5cm}


\end{abstract}


\maketitle

\section{Introduction}
It has been thought for a long while that Cooper
pairs in the Bardeen-Cooper-Schrieffer (BCS)  theory \cite{bcs} with the
conventional
 electron-(acoustic) phonon interaction  are singlets and
their wave function is isotropic (s-wave). This
interaction has been thought to be local
 in space, so it could not
 lead to a higher angular momentum pairing. The pairing symmetry breaking is a many-body effect in accordance
with a well-known quantum mechanics theorem, which
states that the coordinate wave function of  two particles does not
become zero (or has no nodes) in the ground state \cite{landau}. Hence any
superconductor should seem to be s-wave
 in the strong-coupling  limit, where  pairs are  individual (e.g. bipolarons
\cite{alebook,alemot})  rather than overlapping  Cooper pairs.

Recently I have revised the symmetry of the superconducting  state mediated by
 conventional acoustic phonons  \cite{dwave}. The sound speed anisotropy leads to a non-local attraction between carriers and unconventional Cooper pairs
in the BCS  layered superconductors in
a wide range of carrier densities. The  Bose-Einstein condensate (BEC)
  can  also break the rotational symmetry  due to a reduced Coulomb repulsion between unconventional small bipolarons  bound by strong EPI.

Earlier we have proposed that  a
strong departure of the cuprate superconductors from conventional BCS/ Fermi-liquids originates in
the Fr\"ohlich EPI of the order of 1 eV
\cite{ale96,alebra,alebook}, routinely neglected in the Hubbard $U$ and
$t-J$ models of cuprate superconductors
 \cite{band}. This interaction with $c-$axis polarized  phonons
is virtually unscreened because the upper limit for the out-of-plane
plasmon frequency ($\lesssim 200$ cm$^{-1}$ \cite{plasma}) in
cuprates is well below the characteristic frequency of optical
phonons, $\omega_0\approx$ 400 - 1000 cm $^{-1}$. Since
 screening is poor, the magnetic interaction remains
small compared with the Fr\"ohlich EPI at any doping of cuprates (see also Ref.\cite{condmat}). Taking into account that
the direct Coulomb repulsion is of the same order as the Fr\"ohlich EPI,  we have proposed a so-called  Coulomb-Fr\"ohlich model (CFM) of cuprate superconductors with the ground
state in the form of mobile small bipolarons or polaronic Cooper pairs (depending on doping)   \cite{ale96,alekor2,alenar}, which can condense at  high temperatures \cite{jim}. More recently we have  shown that even a weak long-range EPI combined with the
Hubbard $U$ provides sizable superconducting order in doped
Mott-Hubbard insulators and/or strongly-correlated metals \cite{tom}.

Now compelling experimental evidence for a strong EPI
has arrived from isotope effects \cite{zhao}, high resolution angle
resolved photoemission spectroscopies (ARPES) \cite{lanzara}, a
number of optical \cite{mic}, neutron-scattering
\cite{ega,rez}  and some other spectroscopies of cuprates
\cite{davis}, in particular from  recent pump-probe experiments \cite{boz}. These experimental observations and our theoretical findings tell us that EPI with
conventional  phonons  is responsible for the unconventional pairing in cuprate superconductors.

\section{Unconventional Cooper pairs glued by acoustic phonons}
It has
gone unquestioned that the unconventional pairing
 requires unconventional electron-phonon
 interactions  with specific optical phonons
 \cite{abr,sig,shen,kulic,tsai},
  sometimes combined with anti-ferromagnetic fluctuations \cite{mazin} or  vertex corrections \cite{hag},
 or  non-phononic
 mechanisms of pairing (e.g. superexchange \cite{band}), and a specific shape of the Fermi surface. Here I show that even  conventional acoustic phonons can bound  carriers into unconventional Cooper pairs  due to the sound-speed anisotropy in layered crystals.

In the framework of the BCS theory the symmetry of the order
parameter $\Delta(\bf k)$ and the critical temperature, $T_c$, are
found by solving the linearised "master" equation,
\begin{equation}
\Delta({\bf k})=-\sum_{\bf k^{\prime }}V({\bf k,k^{\prime}}){\frac{\Delta ({\bf k^{\prime} })}{2\xi _{{\bf %
k^{\prime }}}}}\tanh \left({\xi _{\bf %
k^{\prime }}\over{2k_BT_c}}\right). \label{master}
\end{equation}
The interaction $V(\bf k,k^{\prime })$ comprises the attraction, $-
V_{ph}({\bf q})$, mediated by  phonons, and the Coulomb repulsion, $V_{c}({\bf
q})$ as
$V({\bf k,k^{\prime}})=-V_{ph}({\bf q})\Theta(\omega_D-|\xi _{\bf
k}|)\Theta(\omega_D-|\xi _{\bf k^{\prime}}|)+V_{c}({\bf
q})\Theta(\omega_p-|\xi _{\bf k}|)\Theta(\omega_p-|\xi _{\bf
k^{\prime}}|)$,
where $V_{ph}({\bf q})=C^2/NMc_l^2$ is  the square of the matrix
element of the deformation potential, divided
by the square of the acoustic phonon frequency, $\omega_{\bf q}=c_l
q$, $c_l$ is the sound speed, $M$ is the ion mass,  $N$ is the
number of unit cells in the crystal, and $\xi _{\bf k}$ is the
electron energy relative to  the Fermi energy.  The
magnitude of $C$ is roughly the electron bandwidth in rigid metallic or semiconducting  lattices. The
electron momentum transfer ${\bf q}={\bf k -k^{\prime}}$ or its
in-plane component has the magnitude $q= 2^{1/2}k_{F}[1-\cos \psi
]^{1/2}$ for the spherical or cylindrical Fermi surface,
respectively, where $\psi $ is the angle between ${\bf k}$ and ${\bf
k^{\prime }}$, and $\hbar k_F$ is the Fermi momentum. Theta
functions  account for a difference in frequency scales of the
electron-phonon interaction, $\omega_D$, and the Coulomb repulsion,
$\omega_p\gg \omega_D$, where $\omega_D$ and $\omega_p$ are the
maximum  phonon and plasmon energies, respectively.

If one neglects anisotropic effects, replacing $V_{ph}({\bf q})$ and
$V_{c}({\bf q})$ by their Fermi-surface averages, $V_{ph}({\bf
q})\Rightarrow V_{ph}$, $V_{c}({\bf q})\Rightarrow V_c$,
then there is only an $s$-wave  solution of Eq.(\ref{master}),
$\Delta_s$, independent of ${\bf k}$.  The sound speed anisotropy
actually changes the symmetry of the BCS state. While $c_l$ is a
constant in the isotropic medium, it depends on the direction of
${\bf q}$  in any crystal. The anisotropy is particulary large in
layered crystals like cuprate superconductors.  As an example,  the measured velocity of longitudinal
ultrasonic waves along $a-b$ plane, $c_{\parallel}$=4370 ms$^{-1}$
 is almost twice larger than that along $c$ axis, $c_{\perp}$=2670 ms$^{-1}$
in Bi$_2$Sr$_2$CaCu$_2$O$_{8+y}$ \cite{chang}. It makes $V_{ph}({\bf
q})$ anisotropic,
\begin{equation}
V_{ph}({\bf q})={C^2\over{NMc_{\perp}^2(1+\alpha
q_{\parallel}^2/q^2)}}, \label{ph}
\end{equation}
where $\alpha=(c_{\parallel}^2-c_{\perp}^2)/c_{\perp}^2$ is the
anisotropy coefficient, which is about $2$ in  cuprates.
The corresponding real-space  potential is non-local,
\begin{equation}
V({\bf
r})= -V_{ph}\Omega \left[{\delta({\bf r})\over{d}}+{\alpha\over{4\pi
(1+\alpha)^{1/2}r^3}}\right],
\end{equation}
 falling as $1/r^3$ at
the distance $r \gg d$ between two carriers in the plane, where
$V_{ph}=C^2/Mc_{\perp}^2$. Also the
Coulomb repulsion is $q$ dependent, $ V_c({\bf q})=4\pi
e^2/V\epsilon_0(q^2+q_s^2). $  In the framework of the random phase
approximation the inverse screening radius squared is found as
$q_s^2=8\pi e^2N(0)/V\epsilon_0$ with the density of states (per
spin), $N(0)$, at the Fermi surface. Here $d$ is the inter-layer distance and  $\epsilon_0$ is the
(\emph{in-plane}) static dielectric  constant of the host cuprate
lattice of the volume $V$.

Solving the master equation (\ref{master}) with  2D electron
spectrum one can expand $\Delta({\bf k})=\sum_m \Delta_m
\exp(im\phi)$ and $V_{ph,c}({\bf q})=\sum_{m}V_{ph,c}(
q_{\perp},m)\exp[im(\phi-\phi^{\prime})]$ in series of the
eigenfunctions of the $c$-axis component of the orbital angular
momentum, where $\phi$ and $\phi^{\prime}$ are  polar angles of the
in-plane momenta, ${\bf k}_{\parallel}$ and ${\bf
k}^{\prime}_{\parallel}$, respectively. The critical temperature of an  $m$-pairing channel ($m=0,\pm 1,\pm 2,...$) is found as
\begin{equation}
T_{cm}=1.14\omega_D\exp \left[-1/(\lambda_m-\mu_m^\ast)\right],
\end{equation}
where $\mu _{m}^{\ast }=\mu _{m}/[1+\mu _{m}\ln (\omega _{p}/\omega
_{D})]$. Here $\lambda_m$ and $\mu_{m}$ are the phonon-mediated attraction
and the Coulomb pseudopotential in the $m$-pairing channel, given
respectively by
\begin{equation} \label{lambda}
{\lambda_m\over{\lambda}}= \delta_{m,0}+{
\alpha\over{2\sqrt{\gamma}}}
\int_0^{\gamma}{dx[x+1-\sqrt{x(x+2)}]^m\over{\sqrt{x+2}}} ,
\end{equation}
and
\begin{equation}
{\mu_{m}\over{\mu_c}}= {\sqrt{\tilde{\gamma}}\over{2}}
\int_0^{\tilde{\gamma}}{dx[x+\beta+1-\sqrt{(x+\beta)(x+\beta+2)}]^m\over{\sqrt{x(x+\beta)(x+\beta+2)}}}, \label{mu}
\end{equation}
where $\lambda= N(0)C^2/NMc_{\parallel}^2$,
$\gamma=\pi^2/2d^2k_F^2(1+\alpha)$,
$\tilde{\gamma}=\gamma(1+\alpha)$, $\mu_c=4e^2d^2N(0)/\pi
V\epsilon_0$, and $\beta=q_s^2/2k_F^2$ (note that $\lambda$,
$\mu_c$, and $q_s$ do not depend on the carrier density since $N(0)$
is roughly constant in the quasi-two dimensional Fermi gas).

The effective attraction of two electrons in the Cooper pair with
non-zero orbital momentum turns out finite at any finite anisotropy,
$\alpha \neq 0$, but numerically smaller than in the $s$-channel, as   seen from its analytical expression for
$s$-wave, $m=0$, $p$-wave, $m=1$,  $d$-wave, $m=2$, and higher orbital momentum pair states,
obtained by integrating in Eq.(\ref{lambda}). The Coulomb repulsion turns out \emph{much} smaller in the unconventional pairing
states than in the conventional $s$-wave state,  which is
seen from the analytical expression for $\mu_m$, Eq.(\ref{mu}) and from Fig.1.

Using the simplest parabolic approximation for the 2D-electron energy
spectrum we can draw some conclusions on the carrier-density
evolution of the order-parameter symmetry. Within this
approximation, $k_F^2=2\pi d n$ and $N(0)= m^\star V/2\pi d \hbar^2$, where
$n=2x/\Omega$ is the carrier density, $m^\star$ is the effective mass, and $x$ is the doping level as
in La$_{2-x}$Sr$_x$CuO$_4$ with the unit cell volume $\Omega$. The
ratio of the parameters $\beta=m^\star e^2\Omega/2\pi \hbar^2 d^2
\epsilon_0 x$ and $\tilde{\gamma}=\pi\Omega/8 d^3x\thickapprox
0.044/x$   is independent of the carrier density,
$\beta/\tilde{\gamma}=4m^\star e^2d/\pi^2\hbar^2\epsilon_0$, which is
approximately $5$ for the  values of $m^\star =4m_e$ and $\epsilon_0=10$.
Fixing the value of the EPI  constant at $\lambda=\mu_c/12$ (which
corresponds to  the weak-coupling BCS regime with $\lambda \approx
0.1$ since $\mu_c$ is of the order of $1$) and taking
$\mu_c\ln(\omega_p/\omega_D)=3$, we draw the \emph{anisotropy-doping
} phase diagram, Fig.2, with the critical lines for $s$, $p$ and $d$
order parameters, defined by $\lambda_m=\mu_m^\ast$. The state with
the lowest magnitude of the anisotropy, $\alpha/(1+\alpha)^{1/2}$,
is physically realized since it has the highest $T_c$. At
substantial doping the screening length becomes larger than the
typical wavelength of electrons, $\beta\rightarrow 0$, so that the
$s$-wave state is the ground state at a large number of carriers per
unit cell for any anisotropy. On the contrary, the Coulomb repulsion
is reduced to the local interaction at a low doping, $
\beta\rightarrow \infty$,  and $d$-wave Cooper pairs are the ground
state even at very low value of the anisotropy, Fig.2.
Interestingly, $s-$ and $d$-states turn out degenerate at some
intermediate value of doping, $x=x_c$. Hence there is a quantum
phase transition with increasing doping from $d-$ to
$s$-superconducting  state, if $\alpha
> \alpha_c$, and from $d-$ to the normal state and then to the $s$-wave
superconductor, if  $\alpha < \alpha_c$,  Fig.2.

\section{Breakdown of rotational symmetry in the strong-coupling limit}

 In the  strong-coupling regime, $\lambda \gtrsim 1$, the pairing is
individual \cite{alebook},   in contrast with the collective
 Cooper pairing. While BEC of individual bipolarons can break
the symmetry  on a discreet lattice \cite{alesym,andsym},  I have
proposed a symmetry breaking mechanism \cite{dwave}, which works even  in a
continuum model, where the ground state, it would seem, be s-wave
 to satisfy the theorem. The unscreened
 Fr\"ohlich EPI  in layered ionic lattices like cuprates
has been suggested by us as the key for pairing \cite{alebook}.
Acting alone it cannot overcome the direct Coulomb repulsion, but
almost nullifies it since $\epsilon_0 \gg 1$. That allows the weaker
deformation potential, Eq.(\ref{ph}), to bind carriers into
real-space bipolarons, if $\lambda \geqslant 0.5$ \cite{alebook}.
While its local part  is negated by  the
strong on-site repulsion $U$, the  non-local tail provides
bound pairs of different symmetries with the binding energies
$\Delta_s > \Delta_p > \Delta_d >...$ in agreement with the theorem.
However, there is the residual Coulomb repulsion between bipolarons,
$v_c(R)$.  Since bipolarons have a finite
extension, $\xi$, there are corrections to the Coulomb law. The
bipolaron has no dipole moment, hence  the most important correction
at large distances between two bipolarons, $R\gg\xi$, comes from the
charge-quadrupole interaction,
\begin{equation}
v_c(R)=4e^2{1\pm
\eta\xi^2/R^2\over{\epsilon_0 R}},
\end{equation}
  where $\eta$ is a number of the order
of 1, and plus/minus signs  correspond to bipolarons in the same or
different planes, respectively. The dielectric screening,
$\epsilon_0$ is highly anisotropic in cuprates, where the in-plane
dielectric constant, $\epsilon_{0\parallel}$, is much larger then
the out-of-plane one, $\epsilon_{0\perp}$ \cite{dielectric}. Hence
the inter-plane repulsion provides the major contribution to the
condensation energy. Since $\xi^2 \propto 1/\Delta$, the repulsion
of unconventional bipolarons with smaller binding energies,
$\Delta_d, \Delta_p < \Delta_s$, is reduced compared with the
repulsion of $s$-wave bipolarons. As a result, with increasing
carrier density we anticipate a transition from  BEC of $s$-wave
bipolarons to BEC of more extended $p-$ and $d$-wave real-space
pairs in the strong-coupling limit.

\section{Conclusions}

Several authors \cite{band} have remarked that superexchange, and
not phonons is responsible for the symmetry breaking in
unconventional superconductors like doped cuprates. Here I arrive at
the opposite conclusion. Indeed, superexchange interaction, $J$, is
proportional to the electron hopping integral, $t$,  squared divided
by the on-site Coulomb repulsion (Hubbard $U$), $J=4t^2/U$,
estimated as $J \thickapprox 0.15$ eV in  cuprates \cite{band}. This
should be compared with the acoustic-phonon pairing interaction,
$V_{ph}$, which is roughly the Fermi energy, $V_{ph}\thickapprox
E_F\thickapprox 4t$ in a metal, or the bandwidth
squared divided by the $ion-ion$ interaction energy of the order of
the \emph{nearest-neighbour} Coulomb repulsion, $Mc_l^2 \thickapprox
V_c$ in a doped insulator. The small ratio of two
interactions, $J/V_{ph}\thickapprox t/U \ll 1$, or
$J/V_{ph}\thickapprox V_c/U \ll 1$ and the giant sound-speed
anisotropy \cite{chang} favor conventional EPI  as the
origin of the unconventional pairing both  in underdoped cuprates,
where the pairing is individual \cite{alebook}, and also in overdoped
samples apparently with  polaronic Cooper pairs, which can coexist with bipolarons \cite{alenar}.

Moreover recent studies by Aimi and Imada~\cite{imada} of the Hubbard model, using a
sign-problem-free  variational Monte Carlo (VMC) algorithm, have shown that previous approximations  overestimated the normal state
energy and
 therefore overestimated the condensation energy by several
 orders of magnitude, so that the Hubbard model does not account
for high-temperature superconductivity. This remarkable
result is in line with earlier numerical studies using the
auxiliary-field quantum (AFQMC) \cite{afqmc} and constrained-path
(CPMC) \cite{cpmc} Monte-Carlo methods, none of which found
superconductivity in the Hubbard model. On the other hand using VMC method we have found that even a relatively
weak finite-range EPI induces the d-wave
superconducting state of doped Mott-Hubbard insulators and/or
strongly-correlated metals with a sufficient condensation energy  \cite{tom}.

 I conclude that  the finite-range
electron-phonon interaction is the key to the high and a higher temperature superconductivity.

I would like to thank Tom Hardy,  Jim Hague, Victor Kabanov, Pavel Kornilovitch and John Samson   for long-standing
collaboration and illuminating discussions.
The work was supported by EPSRC (UK) (grant numbers EP/C518365/1 and EP/D07777X/1).

\newpage
\begin{figure}
\begin{center}
\includegraphics[angle=-90,width=0.95\textwidth]{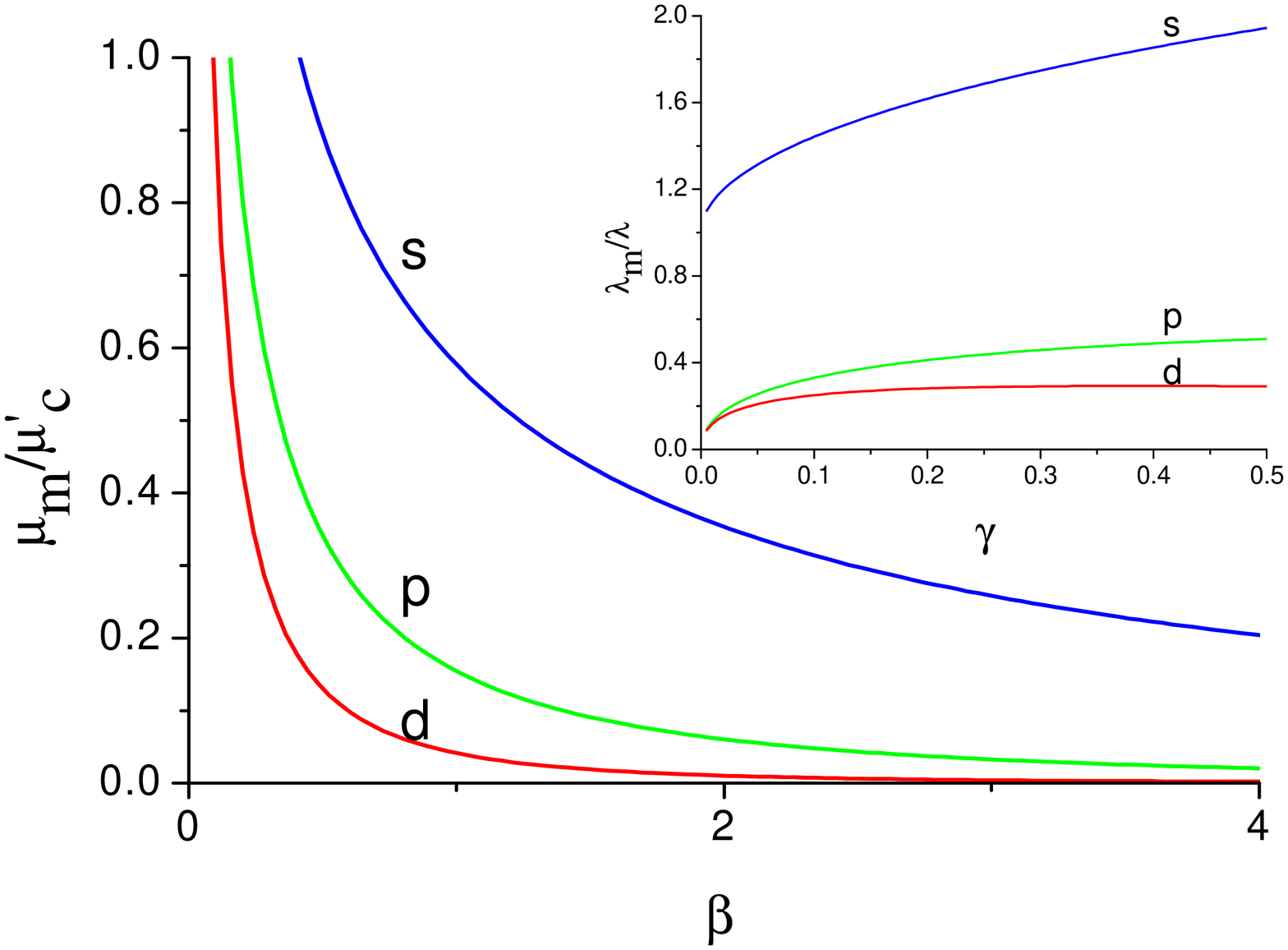}
\vskip -0.5mm \caption{The Coulomb repulsion, $\mu_{m}$, as a
function of the ratio of the electron wavelength to the screening
length squared ($\beta=q_S^2/2k_F^2$), and the electron-phonon
coupling constant, $\lambda_m$,  as a function of the ratio of the
electron wavelength to the inter-plane distance squared,
$\gamma=\pi^2/2d^2k_F^2(1+\alpha)$ for $\alpha=4$ (inset) in $s,p$
and $d$ pairing channels. Here $\mu_c^{\prime}=\mu_c
\tilde{\gamma}$. ({Reprinted with permission from}
Ref.\cite{dwave}.
{\copyright 2008 by the American Physical Society}.)}
\end{center}
\end{figure}

\begin{figure}
\begin{center}
\includegraphics[angle=-90,width=0.95\textwidth]{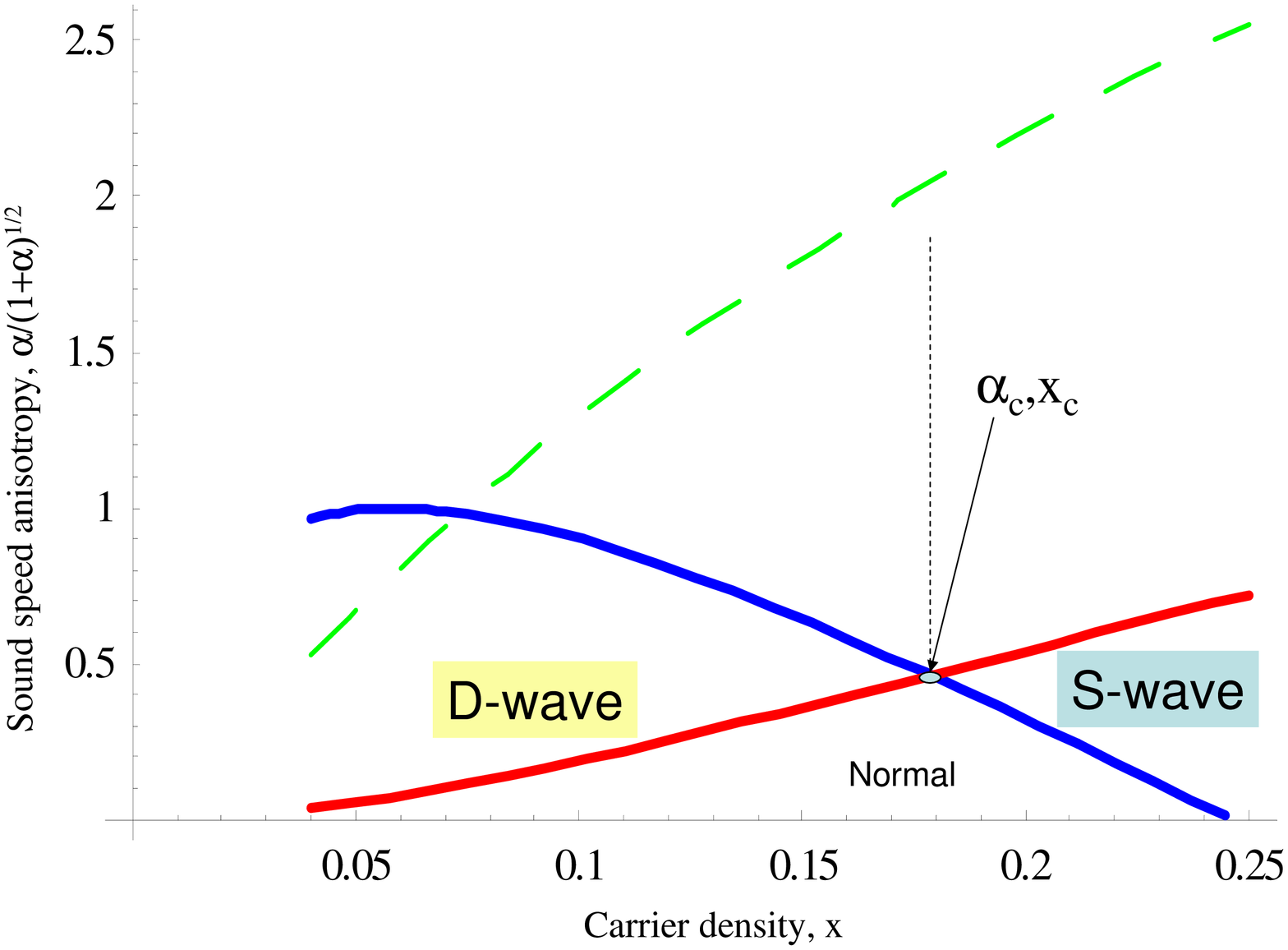}
\vskip -0.5mm \caption{Critical sound-speed anisotropy,
$\alpha/(1+\alpha)^{1/2}=(c_{\parallel}^2-c_{\perp}^2)/c_{\parallel}c_{\perp}$,
as a function of doping, $x$, for $\lambda=\mu_c/12$ (solid lines
correspond to $d$ and $s$ states, and dashed line to $p$-state).
With increasing carrier density there is a quantum phase transition
at $x=x_c$ from a d-wave to an s-wave superconductor, when $\alpha
>\alpha_c$, and  two quantum phase transitions from $d$-wave to
the normal state and from the normal state to the $s$-wave state
when $\alpha < \alpha_c$. ({Reprinted with permission from}
Ref.\cite{dwave}.
{\copyright 2008 by the American Physical Society}.)}
\end{center}
\end{figure}

\end{document}